# Investigations on the improved cycling stability of Kazakhstanite phase Fe-V-O layered oxide by using superconcentrated electrolytes: Generalized solubility limit approach (Part I)

*Arijit Mitra[a], Advait Gilankar[a], Sambedan Jena[b], Debasish Das[b], Subhasish B. Majumder[c], Siddhartha Das[a]\**

[a]Department of Metallurgical and Materials Engineering, Indian Institute of Technology Kharagpur, Kharagpur, West Bengal, India – 721302

[b]School of Nano Science and Technology, Indian Institute of Technology Kharagpur, Kharagpur, West Bengal, India - 721302

[c]Materials Science Center, Indian Institute of Technology Kharagpur, Kharagpur, West Bengal, India – 721302

**Corresponding Author**

 \* Email: sdas@metal.iitkgp.ernet.in; Phone : +91-3222-283256

**ABSTRACT**

In this article, we address the issue of vanadium dissolution pertinent in the layered $Fe_5V_{15}O_{39}(OH)_9.9H_2O$ using the solubility limit approach. This layered oxide is prepared via a low-cost solution phase synthesis route and crystallizes in the Kazakhstanite phase (Space Group: C2/m), confirmed using selected area electron diffraction and x-ray diffraction. The layered oxide exhibits the 2 electron redox reaction of vanadium (V5+ to V3+) along with the 1 electron redox reaction of iron within the voltage window of 1.5-3.8V. This results in a high specific capacity of ~350mAhg$^{-1}$ which can be extracted from this material. However, the transition from $V^{4+}$ to $V^{3+}$ is identified to initiate a dissolution process at ~2.5V, resulting in a loss of active material and poor cycling stability. The vanadium dissolution is found to be arrested by switching to a superconcentrated electrolyte, wherein the amount of "free" solvent is low. An electrolyte, consisting of seven molar lithium bis(trifluoromethanesulfonyl)imide in 1,3-Dioxolane: 1,2-Dimethoxyethane = 1:1 (v:v), is found to be suitable in providing the best cycling stability amongst the other compositions tested. The electrochemical characteristics of the passivation layers formed over lithium foil are mathematically modeled to indicate the preference of superconcentrated electrolytes over relatively dilute ones.

## 1. INTRODUCTION

Lithium-ion rechargeable batteries are undoubtedly the most popular choice of electrochemical energy storage devices available in the market, in recent times. The flexibility in the electrochemical properties, offered by the existing materials chemistry within these batteries, enable them to be used in a variety of applications such as portable electronics, electric vehicles, micro-grid energy storage systems *etc.*[1-4] 2019 witnessed the Nobel Prizes in Chemistry being awarded to the inventors of the lithium-ion batteries.

In-spite of these batteries being the most popular choice for the above mentioned applications, improvements in their properties are still desired. Scientists across the globe are improving the various metrics defining their performance by tuning and varying the materials chemistry. One of the several strategies, which scientists have devised to improve the energy density of these batteries, is to choose a cathode material which has high specific capacity (since energy density equals the area under the V-Q curve). For this purpose, vanadium and molybdenum based materials have been investigated for a long time. They have the potential to exhibit multi-electron redox reactions, which enables them to attain high specific capacities.[5-10] Furthermore, their nominal voltage is also not very low (~2.5V) so as to affect the energy density, or cause any parasitic reactions such as electrolytic decomposition.[7] For example, $V_2O_5$ has a high theoretical specific capacity of 294 mAhg$^{-1}$ for 2 Li$^+$ ion intercalations, which occurs between 2.0-4.0V.[7]

The performance shown by the vanadium based compounds as cathodes for lithium-ion batteries is severely limited by their cycling stability.[11, 12] Over the years, several data have been reported revealing the various mechanisms of lithium insertion into vanadium oxides.[12-14] Delmas et al. have studied the various phase transitions occurring in the $\omega$-$V_2O_5$ when lithium ions are inserted.[15] Their prime conclusion is that the original $V_2O_5$ layered structure is preserved till the first lithium-ion is inserted. On further lithium insertion, there is a formation of an irreversible phase. This irreversible phase formation is reported as one of the causes for poor cycling stability and large specific capacity drops. Similar phase transitions have also been reported for various polymorphs such as $V_6O_{13}$, $LiV_3O_8$, *etc.*[16-18] In some of the texts, it is reported that the vanadium dissolution is prominent similar to the one observed in $LiMn_2O_4$, leading to loss of active material and poor cycling behavior.[19-21] Some of the strategies which have improved the performance of these vanadium containing materials in terms of cycling stability include use of several dopants into the vanadium oxide structure such as Fe, or forming xerogel structures using structural water as a stent to prevent the layered structure from collapsing during operation.[21-24] In spite of such a vast literature available today, the problem of cycling stability still persists in almost all the known vanadium containing materials. The cause for this seems to be material specific.

In this article, we address the issue of cycling stability of one of the vanadium containing cathode material using the solubility limit approach. The material under study is a naturally found mineral phase called Kazakhstanite, which is found in parts of Kazakhstan and Nevada valley. The chemical formula of this phase is $Fe_5^{3+}V_3^{4+}V_{12}^{5+}O_{39}(OH)_9.9H_2O$, as per literature.[25] This phase can deliver a high theoretical capacity compared to other vanadium containing materials, due to the presence of an additional $Fe^{3+}/Fe^{2+}$ redox couple along with the $2e^-$ vanadium redox. In our preliminary electrochemical studies which are discussed later, we identify that the material indeed delivers a very high specific capacity of ~350 mAhg$^{-1}$ within a voltage window of 1.5V-3.8V. However, it suffers from poor cycling stability like other vanadium containing compounds. The energy density which can be extracted from this material ($\sim$800Whkg$^{-1}$) is higher than the ones reported for few commercial cathode materials such as LMO ($\sim$450Whkg$^{-1}$), and comparable to NCM ($\sim$675Whkg$^{-1}$) and NCA ($\sim$785Whkg$^{-1}$). The mechanism and underlying cause for such poor cycling stability is studied using XRD (for possible phase transitions) and XPS (for oxidation state changes) at different states of charge. The cause for the poor cycling stability is identified to be vanadium dissolution from the active material. FMEA (Failure Mode Effect Analysis) reveal that active material dissolution along with continuous irreversible phase transitions must be responsible for the poor long-term cycling behavior of a majority of the cathode materials for lithium-ion batteries, including this Kazakhstanite phase (Supporting information). With a number of commercial cathode materials such as LMO exhibiting long term capacity degradation due to active material dissolution, many researchers have started investigating possible ways to understand and eliminate this issue.[26] We propose the solubility limit concept of dissolving a certain material in a solvent to arrest this vanadium dissolution and tackle the problem. The vanadium species, which dissolves into the Li-ion electrolyte during cycling, should possess a solubility limit with respect to the solvent used in the electrolyte. Once the concentration of the vanadium species reaches its solubility limit, it will not dissolve any further in the electrolyte. This should lead to a stable specific capacity upon further cycling. In order to achieve this solubility limit faster during cycling, one approach can be to reduce the amount of solvent present in the electrolyte. The low of amount of solvent in the electrolyte will change the activity coefficients of the solvated ions due to a different type of solvation shell present. This is expected to reduce the total amount of dissolved vanadium species from the material by not favoring the forward dissolution process. The solubility limit approach is expected to be different from other methodologies adopted to improve the cycleability of cathode materials. This approach attacks the issue of active material dissolution from the perspective of engineering the electrode-electrolyte interface, rather than engineering the electrode material itself. The poor cycling stability of the Kazakhstanite phase due to vanadium dissolution makes this material highly suitable for validating this strategy, since the impact in the cycling stability will be directly contributed by the use of this strategy and will be less prone to statistical deviations related to unintentional processing/fabrication errors. The cycling stability in the layered Fe-V-O Kazakhstanite phase is found to be significantly improved by the use of this approach, with good rate performance. In this manuscript, the improved electrochemical performance of the Kazakhstanite phase is described from the perspective of a reduced elemental dissolution process along with an improved electrochemical performance of the lithium metal anode. This work sets a premise for the generalized solubility limit approach, wherein we hypothesize that this concept can be utilized to improve the cycleability of a variety of vanadium based cathode materials and establish it as a generalized and simple technique to improve the cycleability of vanadium containing cathode materials.

## 2. EXPERIMENTAL SECTION

Iron (III) nitrate nonahydrate (ACS Grade), polyvinyl alcohol, polymethyl methacrylate were procured from Merck. Dimethyl carbonate (>99%) was procured from Loba Chemie. Ammonium metavanadate (ACS Grade), $LiPF_6$ (Battery Grade), $LiClO_4$ (Battery Grade), 1,3-Dioxolane(Reagent Plus), 1,2-Dimethoxyethane (Reagent Plus), ethylene carbonate (98%) and commercial electrolyte were procured from Sigma Aldrich. Lithium Bis(Trifluoromethanesulfonyl) imide (>98%), fluoroethylene carbonate (FEC) (98%) were procured from Tokyo Chemicals Industry. All chemicals were used as received without further purification. All solutions for materials synthesis were made using deionized water.

### 2.1. Synthesis of Layered Fe-V-O Kazakhstanite:

In a 500 ml Borosil Reagent Bottle, 6 mmol of Ammonium Metavanadate was dissolved in 90 ml of water at $100^{o}C$ on a digital hot plate cum stirring unit. In a separate beaker, 2 mmol of Iron (III) Nitrate was dissolved in 10 ml of water. The iron nitrate solution was added dropwise to the ammonium metavanadate solution under vigorous stirring to obtain yellow coloured suspension. The lid of the reagent bottle was closed and the hot plate temperature was raised to $120^{o}C$ with stirring. The yellow suspension was kept under this condition for 4 hrs to obtain a brown coloured suspension. The sediments were filtered and washed several times using deionized water, and finally with acetone, to remove the contaminants. The desired powder was obtained after drying the solid products in a vacuum oven at $60^{o}C$.

### 2.2. Structural Characterization:

The as-prepared powders were structurally characterized using x-ray diffraction and electron microscopy techniques. X-ray diffractogram was obtained using a Bruker D8 Discover Diffractometer (equipped with sample alignment system using laser focusing) with Cu Kα radiation (λ=0.15418nm). Another x-ray diffractogram was collected at the BL-12 Beamline, Indus-2 Synchrotron at a wavelength of 0.82463Å for better resolution of the diffraction peaks. The Pawley Refinement of the obtained diffractogram was performed using GSAS-II Suite.[27] The Rietveld Refinement of the obtained diffractogram was performed using GSAS-II Suite.

High resolution transmission electron micrographs, selected area diffraction patterns, x-ray elemental maps, along with bright and dark field STEM images were obtained using JEOL2100F TEM, operating at 200kV. Image processing of the micrographs was performed using ImageJ software.[28] Scanning electron microscopy was performed using Zeiss Gemini 500 microscope, which includes the EDS spectroscopy of dissolution products on cycled separators from different electrolytes using EDAX Elect Plus detector. Surface topography images along with surface potential maps of the particles were obtained in an atomic force microscope (Agilent 5500AFM), using a PPP-EFM Probe (Nanosensors). FTIR spectrum was collected from a range of 450 cm$^{-1}$ to 4000 cm$^{-1}$ using Nicolet 6700 (Thermo Fischer Scientific). Raman spectroscopy of the as-prepared powder was carried out using T64000 RAMAN spectrometer (Horiba) with Argon-Krypton mixed ion gas laser as excitation source having a wavelength of 532nm. Chemical oxidation states of the elements were identified using Thermo K-Alpha$^{+}$ x-ray photoelectron spectroscope (Thermo scientific). The best resolution of the x-ray photoelectron spectroscope is 0.5eV FWHM at 1eV on Ag 3d peak, with an intensity of 4Mcps. The water content in the as-prepared powders was determined using TG/DTA Analysis performed in Netzsch STA449 under nitrogen atmosphere. The EDS analysis of the cycled separator in preliminary investigations was performed in FEI Inspect F50 scanning electron microscope (FEI), equipped with EDAX Octane Plus energy dispersive x-ray spectroscope.

For ex-situ x-ray diffraction, the x-ray diffractogram was collected using a Bruker D8 Discover Diffractometer with Cu K$_\alpha$ radiation ($\lambda$=0.15418nm), in a 2-Theta Mode. The source angle was fixed at 5$^{\circ}$ for all the cases. The samples were aligned with the goniometer using the laser-based sample alignment apparatus present in the system. Ex-situ XPS was performed using synchrotron x-ray radiation having incident energy of 4.312 keV at INDUS-2 (Beamline-14) Facility. The XPS spectra were recorded using 15 keV Phoibos 225 HV hemispherical analyzer in a fixed analyzer transmission (FAT) mode. The survey scans & core-level scans were measured using step-size of 0.5 eV & 0.1 eV respectively with pass energy of 150 eV. Calibration of the scans was performed by taking the C1S binding energy to be 284.7eV. The peak fitting was performed using Fityk 0.9.8.[29] Raman spectra of the cycled lithium counter electrodes were acquired using Witec Alpha 300R Raman Spectrophotometer (Witec, Germany) using an excitation wavelength of 532nm. The cycled lithium foils were sandwiched between a glass slide and glass cover slip, and sealed using DPX mountant inside the glove box prior to Raman measurements.

### 2.3. Electrochemical Characterization:

A viscous slurry was prepared by grinding the as-prepared powder (60 wt%), acetylene black (20 wt%), polyvinyl alcohol (15 wt%) and polymethyl methacrylate(5 wt%) in N-methyl-2-pyrrolidone in an vacuum mixer (MTI MSK-SFM-7). The slurry was then poured over battery grade Al current collector and tape cast using doctor blade. The coatings were first air dried in hot air oven at 60$^o$C, followed by vacuum drying at 120$^o$C. Circular electrode discs of 15 mm diameter were punched out using a disc cutter (MSK-T06 MTI Corporation, USA). Electrodes with different active material loading, ranging from 1.2 – 3mg (in 15mm discs), were tested. Lithium ion half-cell configuration of CR2032 coin cells were fabricated using the prepared electrodes, with lithium foil as counter and reference electrode. The electrolytes used in the tests were self-prepared inside an argon filled glove box with <0.5ppm for both $H_2O$ and $O_2$ (MBraun Labstar Pro). The various compositions tested are listed in Table 1, along with the separators used during testing. The coin cells were assembled in an argon filled glove box (Mbraun, Germany), with <0.5ppm levels for both $O_2$ and $H_2O$.

| Code | Solvent | LiTFSI Conc. (M) | LiPF6 Conc. (M) | LiClO4 Conc. (M) | Separator Used |
|---|---|---|---|---|---|
| **PL(Commercial)** | EC:DEC = 3:7 | - | 1 | - | Celgard 2400 |
| **EE** | EC:DMC=3:7 with 5 vol% FEC | - | 1 | - | Celgard 2400 |
| **LCL** | EC:DMC = 3:7 with 5 vol% FEC | - | - | 1 | Celgard 2400 |
| **LTLP** | EC:DMC = 3:7 with 5 vol% FEC | 1 | 1 wt% | - | Celgard 2400 |
| **DL1** | DOL:DME = 1:1 | 4 | - | - | Whatman GF/C |
| **DL5** | DOL:DME = 1:1 | 5.5 | - | - | Whatman GF/C |
| **OL** | DOL:DME = 1:1 | 7 | - | - | Whatman GF/C |

**Table 1.** Electrolyte compositions (salt, solvent, concentration) which are tested to improve the cycleability of layered $Fe_5V_{15}O_{39}(OH)_9.9H_2O$

Galvanostatic charge discharge studies were carried out in automated battery testers (BST8-MA, MTI Corporation and BTS4000-5V10mA, Neware) between 1.5V-3.8V vs $Li^+$/Li redox couple. For cycleability studies, the specific current rate was $300mAg^{-1}$. Gamry Series G750 Potentiostat-cum-Galvanostat was used for carrying out electrochemical impedance spectroscopy (EIS) measurements for cycled electrodes after 100 cycles (conducted at 3.8V in the frequency range of 100 kHz to 0.01Hz, with potentiostatic signal amplitude of 5 mV. EIS spectra were analyzed using ZSimpWin 3.21 program.[30] Kramers-Kronig Extrapolation was performed on the impedance spectra for cycled electrochemical cell for a reliable fit of the experimental points with the proposed model.

## 3. RESULTS AND DISCUSSION

### 3.1. Structural aspects of the synthesized material

The structure of Kazakhstanite is not well studied as deserved in the literature. Herein, we have made an attempt to study and understand it.

**Figure 1(a)** shows the X-ray diffractogram of the as-prepared powders. The peaks obtained from the diffractograms match with the Kazakhstanite type phase (PDF # 00-046-1334). The data card does not contain the atomic positions of the elements present in the Kazakhstanite structure. However, the chemical formula is presented as $Fe_5^{3+} V_3^{4+} V_{12}^{5+} O_{39}(OH)_9.9H_2O$. The iron atoms are expected to be in their +3 state, with the vanadium atoms present in both +4 and +5 states. A Pawley refinement of the collected spectrum (Figure 1(b)) provides a good match with the unit cell parameters presented in the data card. The space-group of the phase is identified to be C2/m with the unit cell parameters as a = 11.84 Å, b = 3.66 Å, c = 21.58 Å and β = 98.55$^o$. This is also confirmed by indexing an electron diffraction pattern which is obtained for the as-prepared powder, as shown in Figure 1(c). The detailed calculation performed for indexing the electron diffraction pattern is presented in Supporting Information. Since, the structure is a layered one based on vanadium oxide, we believe that the structure should be very close to that of $V_2O_5$. Indeed, it is observed that that the lattice parameters of the Kazakhstanite phase are nearly integral multiples of the lattice parameters of $V_2O_5$. Thus, the Kazakhstanite phase is expected to be constructed in similar fashion as $V_2O_5$, wherein the unit cell consists of alternating V-O square-pyramidal polyhedral arranged in a layered fashion. The Fe atoms coordinate with the anions probably in the middle of the unit cell. In order to confirm this hypothesis, a vibrational spectroscopy is performed in order to determine the symmetry of coordination between the Fe/V and O atoms. Figures 1(e)-(f) show the Fourier Transform Infrared and Raman spectra of the as-prepared material, from $100cm^{-1} - 1200cm^{-1}$ (Raman) and $400cm^{-1} - 4000cm^{-1}$ (FTIR). Peak deconvolution in the Raman spectra reveals the stretching and bending vibrations of V-O and Fe-O co-ordinations present within the material. The positions for the V-O vibrations are very similar to the ones observed for $V_2O_5$.[31] This further confirms that the synthesized Kazakhstanite phase is made of similar sub-units present in $V_2O_5$. $V_2O_5$ is usually represented as alternating V-O square-pyramidal polyhedral arranged in a layered fashion. We observe, from the Raman spectra, that a similar arrangement should be present in Kazakhstanite. The vibrations observed in FTIR spectrum also confirm the presence of the V-O square- pyramidal sub-units.

The vibrations, indexed to Fe-O bonds in FTIR and Raman Spectra, indicate that they are present in an octahedral co-ordination. Additionally, the presence of O-H bond is also indicated in the FTIR Spectrum. The reported chemical formula in the literature indicates that the phase contains about 9 molecules of water.[25] This is confirmed by performing a thermogravimteric analysis, the results of which is reported in the Supporting information. Thus, the Kazakhstanite phase consists of V-O square pyramidal and Fe-O octahedral units arranged in a layered fashion, with few of the O replaced by OH and $H_2O$. Based on the above findings, a set of atomic positions are created for the V, Fe and O atoms such that above conclusions are satisfied. H atoms are very difficult to map and predict due to their low atomic scattering factor for X-Rays. This set of atomic co-ordinates for Fe, V and O are added to the data card and used for Rietveld Refinement with the experimental diffractogram obtained at the BL-12 Beamline of Indus-2 synchrotron facility. We have used this diffractogram since the one obtained with the lab Cu-K$_\alpha$ is not of sufficient quality for this purpose. As shown in Figure 1(d), the calculated x-ray spectrum matches well with the experimental x-ray spectrum, with R$_{wp}$ of 6.59%. Thus, the atomic co-ordinates deduced from spectroscopic and diffraction experiments can be accepted for future studies in this manuscript. It is to be noted that ab-initio structure determination cannot be performed on the as-prepared samples, since the peaks are found to be broad and not well-resolved (due to microstructural constraints). Even the x-ray diffractogram obtained from the synchrotron beamline BL-12, at Indus-2, is broadened due to microstructural constraints (Supporting Information). Fourier Mapping will also not help in this regard to identify the positions of H atoms. The proposed unit cell for Kazakhstanite phase is presented in Supporting Information.

**Figure 2** shows the scanning and transmission electron micrographs of the as-prepared powders. The morphology of the particles in the as-prepared powders, as shown in Figure 2(a), is flake-type. The unit cell calculated from the diffraction experiments indicate that the two of the dimensions of the unit cell are quite large (a and c). If the size of flakes (along these axes) is not large with respect to the unit cell dimensions, then peak broadening will be observed in the diffractograms. This is very well confirmed from the scanning electron micrographs. The flake-type morphology has one of its dimensions within 100 nm, which is responsible for peak broadening in the x-ray spectrum. It is due to this particular morphology that the ab-initio structure determination from x-ray diffraction is difficult. The peak intensities are not well resolved to perform a charge-flipping technique. The transmission electron micrographs (Figure 2(b)) show that the thickness of the flake is within 20nm (circled region). The fringe spacing along this thickness is measured to be ~0.37nm, which is close to the d-spacing of (0 0 6) plane. The flake appears to be bent in the circled region in Figure 2(b), which has helped us in identifying the thickness. This indicates that the bulk of the flake, outside the circled region in Figure 2(b), should be oriented along [0 0 1] zone axis. This is confirmed by indexing the selected area diffraction pattern shown in Figure 1(c). The selected area diffraction pattern shown in Figure 1(c) was collected from the region of interest shown in Figure 2(b). As indicated in Figure 2(c), a high resolution image in the bulk of the flake shows lattice fringes, which is identified post indexing the FFT spectrum of image. The EDS map of the particle reveals both iron and vanadium to be present uniformly, at an atomic fraction of 1:3.

**Figure 3** shows the core level X-ray photoelectron spectrum of the as-prepared powders, for O, V and Fe. The position of the peaks in Fe $2p^{3/2}$ core level spectrum indicates that it is present in its +3 oxidation state. There are two peaks for $Fe^{3+}$ state observed post deconvolution. This may indicate that Fe is not co-ordinated by similar anions. The water molecule and hydroxide anion must be co-ordinated to iron atoms in the Kazakhstanite phase. From the V$2p^{3/2}$ core level spectrum, it is observed that the vanadium is present in its +4 and +5 state. The ratio of the oxidation states, determined from the area, is close to 1:4, which matches well with the chemical formula reported in the literature. A deconvolution of O1s core level spectrum reveals the presence of three peaks. This is due to the presence of oxygen in three forms, i.e., oxide, hydroxide, and structural water molecule, in the Kazakhstanite phase. From the intensity calculations, it is observed that the Fe and V atoms are present in a ~1:3 atomic concentration ratio, which matches well with the reported chemical formula in the literature.

*3.2.Studies on the issue of vanadium dissolution in the as-prepared powders and mechanism of reversible lithium electrochemistry*

**Figures 4(a) and (b)** show the cycleability curve for the prepared electrodes in a half-cell configuration with EE electrolyte at a specific current of $300 mAg^{-1}$. It is observed that in the initial cycles, a high specific capacity of $\sim 320 mAhg^{-1}$ is obtained, which continuously fades to about half of its initial value in 100 cycles. It is observed that the plateau region at $\sim 2.5$ V disappears rapidly upon cycling, which is the major cause for specific capacity loss.

In order to understand the mechanistic behavior and the possible phase transitions in the material at different states of charge, an ex-situ XRD and XPS is performed at different positions in the derivative plots of specific capacity-voltage curves, as shown in **Figures 5(a), (b) and (c)**. The derivative plot in Figure 5(a) indicates that there are 3 major regions of interest, where oxidation state changes or phase transitions must occur during lithiation behavior. These are OCP-2.85V, 2.85V-2.4V and 1.5V-2.2V. The ex-situ XRD and XPS measurements are performed to identify the behavior of material in these three regions. From the peak position of (0 0 2) planes in the x-ray diffractograms, it is observed that there is a major peak shift to the right when the lithiation behavior starts in OCP-2.85 V region. This indicates that the unit cell contracts along its c-axis upon lithium insertion. The peak positions for the rest of the planes like (1 1 0) do not undergo such drastic changes, as presented in the Supporting Information. The contraction of the c-axis is present for the whole of the plateau-like region from 2.85V-2.4V and continues till 2.2V, beyond which it slightly expands upon further lithium insertion. This led us to believe that the cause for such phenomenon could be release of structural water in the material, which essentially contracts the unit cell along c-axis. This release should be extremely detrimental to the $LiPF_6$ salt, which will cause a drop in the cycleability of the material. On the other hand, oxidation state changes in the material are clearly identified at the regions of interest from ex-situ XPS. From capacity calculations, the number of electrons transferred during lithiation is calculated to be 5, 8 and 12 in the regions OCP-2.85V, 2.4V-2.85V and 1.5V-2.2V, respectively. The above calculations are used to correlate with the results from XPS measurements. In the OCP-2.85V region, the capacity mainly comes from the reduction of iron from +3 to +2 state. In the plateau region between 2.85-2.4V, the capacity comes from the reduction of vanadium to its +4 and +3 states. The +5 state is not completely reduced in this plateau. Vanadium is present in +3, +4, and +5 oxidation states at 2.2V. Since the cycleability loss comes mainly due to the disappearance of the plateau at ~ 2.5V, we believe that the redox reaction of vanadium at this plateau region is somehow being rendered irreversible. We have deduced two possible reasons for this based on the above results. The first possible reason can be electrolyte degradation due to release of structural water, which affects the charge-transfer reaction at the electrode-electrolyte interface. The second possible reason is that the ionic compound formed at the end of the plateau region is soluble in the electrolyte, thereby creating a situation of the loss of active material. In order to determine whether the first case is the cause for the poor cycleability, another cycleability test is conducted with the LCL electrolyte, wherein the $LiClO_4$ salt will not be degraded by the release of water molecules. To our surprise, similar trend is observed in the cycleability plots with LCL electrolyte (shown in Supporting Information). Changing the salt to LiTFSI yielded similar results (Supporting Information). In case of $LiPF_6$, the reaction with the water molecules should have caused a LiF layer to form over the materials and add impedance. However, $LiClO_4$ and LiTFSI are relatively stable salts with water, and shouldn't form any decomposed layer like LiF over the electrode and add impedance. Therefore, it can be safe to assume for the time being that the loss of water molecules may not have much effect on the degradation in the cycling behavior of the material. In order to confirm whether the second case is the cause for the poor cycleability, we have performed a post-mortem analysis on the separator after cycling to check whether any traces of active material are found over it. We have performed an EDS analysis on the separator towards the side facing the lithium foil during cycling in order to check whether the ions have migrated towards lithium foil during cycling. The EDS results, shown in Figure 5(d) and (e), indicate traces of vanadium and iron over the separator towards the side facing the lithium foil. This observation proves that the loss of active material should be the cause for poor cycleability in the material. The dissolution of the active material should occur along the edges of the flake-type particles, which is identified from the KFM maps (Supporting Information). KFM maps reveal that the magnitude of the surface potential is higher in the surface region in comparison to the bulk. We have also noticed that the lithium foil extracted from the cycled cell has its surface turned yellow and black, which might be due to deposition of vanadium and iron compounds. Their ions are introduced into the electrolyte during dissolution process from the material.

The results obtained by us are somewhat in direct contradiction to the electrochemical results published for the Kazakhstanite material in the literature. Wei *et al.* reported the electrochemical performance of the Kazakhstanite phase for the first time, wherein they reported the performance of this material for 2000 cycles without any significant fading.[32] While we do not criticize or question the results published by them, we believe that the results provided by Wei *et al.* do not reveal a clear picture regarding the electrochemical properties of Kazakhstanite phase. Upon careful inspection of the results published, we find that 2000 cycles are completed by cycling the cell at a specific rate of $10Ag^{-1}$. The specific capacity obtained in this condition is $\sim150mAhg^{-1}$. Therefore, each cycle should have taken 0.03hr (both charge and discharge) to complete. Total time taken to complete 2000 cycles is therefore 60hrs, which is 2.5 days. **This timeframe is too short to determine the performance of a material in terms of cycleability.** If the issue of active material dissolution persists, it will not reflect in the electrochemical results since the timeframe of the testing was too short to detect it.

Therefore, the as-prepared Kazakhstanite phase provides a high specific capacity of ~300mAhg$^{-1}$ within a reasonably good working potential of 1.5V-3.8V. This makes it attractive for use as cathode material in lithium-ion batteries. However, the dissolution of active material during cycling imparts it with poor cycleability characteristics.

### 3.3. Proposed solution of solubility limit approach towards improving the cycleability of Kazakhstanite phase

In-spite of possessing a good specific capacity and working potential range, the cycleability of the as-prepared Kazakhstanite phase limits the performance of the material as cathode in lithium-ion batteries. The cause for this poor cycling behavior is found out to be dissolution of one of the intermediate products formed during lithiation. We are proposing a solubility limit approach to circumvent this issue and improve the cycleability.

A material can be dissolved in a certain solvent, whose solubility characteristics are determined by physical and chemical nature of the solute and solvent. A solute will dissolve in the solvent if the free energy of dissolution is negative, which is favored by high entropy of mixing and low enthalpy of dissolution. These parameters are affected by factors like surface tension between the solute particle and solvent, polarity, temperature, complexing behavior of the solvent and solute ion, lattice enthalpy, solvation energy *etc.* However, a material cannot infinitely dissolve in a certain solvent, and possesses a certain solubility limit. The material cannot be dissolved in a solvent beyond this critical value. Since the intermediate product in the Kazakhstanite phase must dissolve in the solvent present in the electrolyte, lowering its concentration should improve the electrochemical characteristics. Further improvement can be made if there are no "free" solvent molecules present in the electrolyte, and all the solvent molecules essentially are complexed to the lithium salt added to it to prepare the electrolyte. The activity of the solvated product can be modified by having less solvent molecules in the solvation shell, which will have a direct impact on the equilibrium constant for dissolution.

Thus, increasing the concentration of lithium salt in a suitable solvent can circumvent the problem of active material dissolution. Of the common lithium salts used in electrolytes, we have preferred using Lithium Bis(Trifluoromethanesulfonyl) imide since it possesses a low lattice energy in order to favor its dissolution at high concentrations.[33] Also, a good choice for solvent to test the proposed solution will be an ether-based solvent mixture such as DOL:DME=1:1, since the solvent molecules form a nanodomain based solvated structure with strong chelating capabilities towards imide based salts.[33-35] Such a chelation will ultimately take the electrolyte to a quasi-solid state.[34]

**Figure 6(a)** shows the cycling performance of the as-prepared material for 100 cycles with different electrolytes mentioned in Table 1, at a specific charge and discharge current of 300 mAg$^{-1}$. The y-axis of this plot is normalized with respect to the specific capacity obtained in the first cycle for each electrolyte. These values are mentioned in the plot. It is observed that the specific capacity converges to a stable value for the concentrated electrolytes after certain number of cycles. The specific capacity continues to deteriorate upon cycling for relatively dilute electrolytes, and drops below 50% for some of the electrolyte compositions. For highly concentrated electrolyte like OL however, the highest obtained specific capacity drops significantly as compared to other compositions in the concentrated domain. DL1 composition, which is 4M LiTFSI in DOL-DME = 1:1, shows a high specific capacity compared to the one observed for OL and DL5. However, the cycleability plot indicates that minor capacity fading is still present in DL1, which is due to the presence of higher concentration of "free" solvent molecules (the molecules not solvating the ionic species in the electrolyte). Since cycleability is an important characteristic for long-term reliability of batteries, OL electrolyte is preferred in comparison to other compositions tested in spite of delivering a lower specific capacity at the tested specific current of $300 mAg^{-1}$. It is to be noted that as the concentration of the salt is increased in the electrolyte, the viscosity is drastically increased as well. This makes the fabrication of the coin cells difficult, since inert gas bubbles might get trapped when the electrodes are sandwiched between the electrolyte. The wave like shape of the cycleability for OL as well is probably due to the high viscosity of the OL electrolyte. Figure 6(b) shows the long term behavior of the electrodes tested with OL electrolyte at a specific charge and discharge current of $100 mAg^{-1}$, for different active material loading. It is observed that the electrode with lower active material loading fades to slightly lower specific capacity value in comparison to one with higher active material loading before having its specific capacity value stabilized. This is expected since the fraction of active material loss due to dissolution will be more for electrode with lower active material loading. However, it is to be noted that the rate of specific capacity loss reduces after certain number of cycles for both cases. The specific capacity drops to ~20% from the value recorded at $60^{th}$ cycles, after 440 cycles (making the total cycles tested as 500). The total time which is recorded for testing 500 cycles is ~2200 hrs, which is close to 3 months.

Figure 6(c) shows the charge-discharge profiles recorded for the electrode with higher active material loading, at various cycles mentioned in the plot. It is observed that the plateau region at ~2.5V still diminishes upon cycling. However, its effect on the cycling performance is less profound. The region from OCP-2.85V hardly shows any capacity fading, which is significant for relatively dilute electrolyte compositions. The charge-discharge profile resembles a sloping line with nearly fixed slope, at the end of 500 cycles.

**Figures 7(a) and (b)** show the rate capability of the electrodes, tested with OL electrolyte, at different specific current ranging from 50mAg$^{-1}$ to 1500mAg$^{-1}$. It is observed that the specific capacity drops in a stepwise manner as the specific current increases. The specific capacity, within each specific current range, does not show any drop or fading. Due to lower ionic conductivity of the concentrated electrolyte, the specific capacity obtained at higher specific currents is low. In-order to simulate a practical scenario of the use of the material in electric vehicles or portable electronics, the electrode is tested with a testing scheme, wherein the electrode is charged rapidly to 3.8V at a specific current of 600mAg$^{-1}$ followed by holding it at 3.8V for 51 minutes. Based on the time recorded to reach 3.8V with a charging specific current of 600mAg$^{-1}$, the total charging time with the above CCCV scheme is ~1 hr. The electrode is then discharged at different specific currents. Figures 7(c) and 7(d) represent the results obtained from the above mentioned testing scheme. It is observed that the discharge specific capacity delivered is increased from the values reported in Figure 7(a) and (b), and is delivered with a high coulombic efficiency. The specific capacity values at different specific current, during discharge, follow a similar trend to the ones observed for the CC charging scheme. There are no distinct differences in the discharge profile for both the cases. Few erroneous points are obtained during charging in some of the cycles. This is identified due to a current spike on the cell when the program shifts from CC charging mode to CV Charging mode. The battery tester shifts the charging voltage to 5V, and then immediately drops to set potential of 3.8V. The raw image of the time-voltage-current plot, recorded in the tester, is presented as supporting information to show the observed current spikes.

The obtained results are compared with some of the reported results in the literature in a tabular format, in Table S1 (Supporting information). With the use of the solubility limit approach, the cycleability is improved. EDS measurements on the separators recovered after cycling indicate that there is a significant amount of Fe and V present in the separators recovered from electrochemical cells cycled with EE electrolyte. On the other hand, we could observe the presence of negligible amount of Fe and V in the separators recovered from electrochemical cells cycled with DL1 electrolyte (Supporting information). The Raman spectrum obtained from cycled lithium counter electrodes indicates the presence of $VO_x$ species in the case of EE electrolyte. Signals marking the presence of vanadium compounds are not observed in the case of cycled lithium counter electrodes cycled with DL1 and OL electrolytes (Supporting Information). This indicates that the amount of dissolved active material in the electrolyte is reduced as one transitions from relatively dilute to superconcentrated electrolytes. It is important to note that the total time taken to complete the cycleability test should be high in order to identify whether any dissolution process is the cause of poor cycleability or not. If the charging/discharging specific current or C rate is high, then the intermediate phase (which dissolves into the electrolyte) will not kinetically form and subsequently, its effect on the cycleability will manifest at a later stage. Therefore, even if a large number of cycles are completed with high charging/discharging rates, the capacity loss due to dissolution process (if present) will not be observed in earlier stages of the cycleability test. The drop in specific capacity will be observed only after a certain number of cycles in such cases.

The concept of super-concentrated electrolytes is fairly recent, where the majority of the research was done past decade.[34-40] A lot of molecular structures have been proposed using MD simulations, in order to understand their electrochemical characteristics.[41, 42] However, their interactions with the different cathode materials are different. We have made an attempt to understand the electrolyte-electrode interface of the Kazakhstanite phase in the half-cell tested using OL electrolyte. Based on the results obtained while studying the mechanism of reversible lithium electrochemistry, it is observed that there is an active material dissolution along with plating of iron and vanadium compounds on the lithium foil during cycling. The newly formed surface film will affect the migration of lithium ions to/from the lithium metal surface during operation. A schematic representation of the electrochemical cell is presented in Figure 7(e). In the electrochemical impedance spectroscopy experiment for this cell, the equivalent impedance will be measured with respect to the $Li^+/Li$ redox reaction occurring at the surface of the lithium foil. Therefore, the impedance contribution of the surface films (both SEI and Fe/V based passivation layer) formed over the lithium foil will be accounted for in the equivalent circuit determination from the impedance spectrum. The lithium ions have to diffuse through these two layers before migrating to the electrolyte phase *via* charge transfer. Since, the thickness of these layers are not infinite with respect to the direction of ionic movement, this becomes a case of linear bounded diffusion. It cannot be considered as a case of restricted diffusion since the diffusing species is not reflected (rather consumed) at the one end of the layer. We present a mathematical derivation for the diffusion of Li-ions through this bilayer. We assume the thickness of the passivation layer to be $L_1$ and thickness of the SEI layer to be $L_2$. Let the diffusivities through the passivation layer be $D_1$, and through the SEI be $D_2$. During the migration of lithium-ions through these layers, it is assumed that there are no changes in their thickness. Also the diffusivities are assumed to be constant throughout the thickness. The transport properties of Li-ion through these layers will follow the Fick's Second law of diffusion, when a small perturbation is introduced over the steady state condition.

$$\frac{\partial c_1}{\partial t} = D_1 \frac{\partial^2 c_1}{\partial x^2} \ (0 < \text{x} < L_1) \ ;$$

$$\frac{\partial c_2}{\partial t} = D_2 \frac{\partial^2 c_2}{\partial x^2} \ (L_1 < \text{x} < L_1 + L_2) \tag{1}$$

Applying Laplace transformation to equation set (1) to convert them into linear differential equations:

$$-c_1(x,0) + s\hat{c}_1(x,s) = D_1 \frac{d^2\hat{c}_1(x,s)}{dx^2} \ ;$$

$$-c_2(x,0) + s\hat{c}_2(x,s) = D_2 \frac{d^2\hat{c}_2(x,s)}{dx^2} \tag{2}$$

where $c_1$(x,0) and $c_2$(x,0) are the steady state initial conditions before the perturbation was introduced. We introduce variables $\Delta c_1$ and $\Delta c_2$, which are the difference between the instantaneous concentration values from their steady state values. Therefore, the equation set (2) can be re-written as:

$$s\Delta\hat{c}_1(x,s) = D_1 \frac{d^2\Delta\hat{c}_1(x,s)}{dx^2} \ ;$$

$$s\Delta\hat{c}_2(x,s) = D_2 \frac{d^2\Delta\hat{c}_2(x,s)}{dx^2} \tag{3}$$

The characteristic solutions for the above linear differential equations are (no particular solution is there since constant = 0):

$$\Delta \hat{c}_1(x,s) = \alpha_1 \exp\left(\sqrt{\frac{s}{D_1}} \, x\right) + \alpha_2 \exp\left(-\sqrt{\frac{s}{D_1}} \, x\right) \; ;$$

$$\Delta \hat{c}_2(x,s) = \beta_1 \exp\left(\sqrt{\frac{s}{D_2}} \, (x - L_1)\right) + \beta_2 \exp\left(-\sqrt{\frac{s}{D_2}} \, (x - L_1)\right) \tag{4}$$

The diffusion flux $\Delta J$ in the Laplace space can be calculated by calculating the derivative of the equation set (4) and multiplying by appropriate pre-factor (-$D_1$ for $\Delta \hat{J}_1$ and –$D_2$ for $\Delta \hat{J}_2$).

$$\Delta \hat{J}_1(x,s) = -\sqrt{sD_1} \left[\alpha_1 \exp\left(\sqrt{\frac{s}{D_1}} \, x\right) - \alpha_2 \exp\left(-\sqrt{\frac{s}{D_1}} \, x\right)\right] \; ;$$

$$\Delta \hat{J}_2(x,s) = -\sqrt{sD_2} \left[\beta_1 \exp\left(\sqrt{\frac{s}{D_2}} \, (x - L_1)\right) - \beta_2 \exp\left(-\sqrt{\frac{s}{D_2}} \, (x - L_1)\right)\right] \tag{5}$$

A matrix formulation can be constructed based on the ideation by Chen *et al.* and Diard *et al.* to represent the concentration and the diffusion flux at a certain point in space, with respect to another point (both points within the boundary).[43, 44]

$$\begin{vmatrix} \Delta \hat{c}_1(x,s) \\ \Delta \hat{J}_1(x,s) \end{vmatrix} = \begin{vmatrix} \cosh\left(\sqrt{\frac{s}{D_1}} \, \delta\right) & \frac{\sinh\left(\sqrt{\frac{s}{D_1}} \delta\right)}{-\sqrt{sD_1}} \\ -\sqrt{sD_1}\sinh\left(\sqrt{\frac{s}{D_1}} \, \delta\right) & \cosh\left(\sqrt{\frac{s}{D_1}} \, \delta\right) \end{vmatrix} \begin{vmatrix} \Delta \hat{c}_1(x + \delta,s) \\ \Delta \hat{J}_1(x + \delta,s) \end{vmatrix} \tag{6}$$

Similarly, the matrix formulation for the region $L_1 < x < L_2$ can be written as

$$\begin{vmatrix} \Delta \hat{c}_2(x,s) \\ \Delta \hat{J}_2(x,s) \end{vmatrix} = \begin{vmatrix} \cosh\left(\sqrt{\frac{s}{D_2}}\,\delta\right) & \frac{\sinh\left(\sqrt{\frac{s}{D_2}}\,\delta\right)}{-\sqrt{sD_2}} \\ -\sqrt{sD_2}\sinh\left(\sqrt{\frac{s}{D_2}}\,\delta\right) & \cosh\left(\sqrt{\frac{s}{D_2}}\,\delta\right) \end{vmatrix} \begin{vmatrix} \Delta \hat{c}_2(x+\delta,s) \\ \Delta \hat{J}_2(x+\delta,s) \end{vmatrix} \qquad (7)$$

Now, the diffusion impedance $Z_{mt}$ is $(\partial E/\partial i)_{mt}$ which can be written as $(\partial E/\partial c)(\partial c/\partial J)(\partial J/\partial i)$ by chain rule. The terms $(\partial E/\partial c)$ and $(\partial J/\partial i)$ can be clubbed as a constant k. These values can be obtained by plugging in the appropriate expressions for dependence of electrochemical potential with concentration (can be calculated by estimating the SOC of the material), and dependence of current density with current. Therefore diffusion impedance is directly proportional to $(\partial c/\partial J)$, which can now be obtained from the equations (6) and (7). For x=0 and $\delta$=L$_1$ (the end points of the passivation layer) in equation (6), and x=L$_1$ and $\delta$=L$_2$ (the end points of the SEI layer) in equation (7), we obtain

$$\begin{vmatrix} \Delta \hat{c}_1(0,s) \\ \Delta \hat{J}_1(0,s) \end{vmatrix} = \begin{vmatrix} \cosh\left(\sqrt{\frac{s}{D_1}}\,L_1\right) & \frac{\sinh\left(\sqrt{\frac{s}{D_1}}\,L_1\right)}{-\sqrt{sD_1}} \\ -\sqrt{sD_1}\sinh\left(\sqrt{\frac{s}{D_1}}\,L_1\right) & \cosh\left(\sqrt{\frac{s}{D_1}}\,L_1\right) \end{vmatrix} \begin{vmatrix} \Delta \hat{c}_1(L_1,s) \\ \Delta \hat{J}_1(L_1,s) \end{vmatrix};$$

$$\begin{vmatrix} \Delta \hat{c}_2(L_1,s) \\ \Delta \hat{J}_2(L_1,s) \end{vmatrix} = \begin{vmatrix} \cosh\left(\sqrt{\frac{s}{D_2}}\,L_2\right) & \frac{\sinh\left(\sqrt{\frac{s}{D_2}}\,L_2\right)}{-\sqrt{sD_2}} \\ -\sqrt{sD_2}\sinh\left(\sqrt{\frac{s}{D_2}}\,L_2\right) & \cosh\left(\sqrt{\frac{s}{D_2}}\,L_2\right) \end{vmatrix} \begin{vmatrix} \Delta \hat{c}_2(L_1+L_2,s) \\ \Delta \hat{J}_2(L_1+L_2,s) \end{vmatrix} \qquad (8)$$

In order to maintain the conditions for continuity and conditions for no charge accumulation at the interface, we have $\Delta \hat{c}_1$ (L$_1$, s) = $\Delta \hat{c}_2$ (L$_1$, s) and $\Delta \hat{J}_1$ (L$_1$, s) = $\Delta \hat{J}_2$ (L$_1$, s). We introduce three variables which are m = -D$_1$/L$_1$, $\Lambda = \sqrt{D_1/D_2}$ and $\lambda = \sqrt{D_1/D_2}$(L$_2$/L$_1$) to simplify the matrix, and represent $\sqrt{s/D_1}\,L_1$ as u. Therefore, the two equations in the set (8) can be combined to obtain

$$\begin{vmatrix} \Delta\hat{c}_1(0,s) \\ \Delta\hat{J}_1(0,s) \end{vmatrix} = \begin{vmatrix} a_{11} & a_{12} \\ a_{21} & a_{22} \end{vmatrix} \begin{vmatrix} \Delta\hat{c}_2(L_1+L_2,s) \\ \Delta\hat{J}_2(L_1+L_2,s) \end{vmatrix} \tag{9}$$

where

$$a_{11} = \cosh(u)\cosh(\lambda u) + \frac{1}{\Lambda}\sinh(u)\sinh(\lambda u)$$

$$a_{12} = \frac{\cosh(u)\sinh(\lambda u)}{-\sqrt{sD_1}} + \frac{\sinh(u)\cosh(\lambda u)}{-\sqrt{sD_2}}$$

$$a_{21} = -\sqrt{sD_1}\sinh(u)\cosh(\lambda u) - \sqrt{sD_2}\cosh(u)\sinh(\lambda u)$$

$$a_{22} = \cosh(u)\cosh(\lambda u) + \Lambda\sinh(u)\sinh(\lambda u)$$

$Z_{mt}$ is directly proportional to $\Delta\hat{c}_1$ (0, s)/ $\Delta\hat{J}_1$ (0, s), which can now be effectively calculated by putting the appropriate expression for $\Delta\hat{c}_2$ and $\Delta\hat{J}_2$ in equation (9) based on the boundary conditions present in the system under consideration. Since lithium foil surface exists at x=$L_1$+$L_2$, it acts as an infinite source of the diffusing lithium ions. Thus, $\Delta\hat{c}_2$ ($L_1$+$L_2$, s) = 0. Therefore, the expression for $Z_{mt}$ can be written as

$$Z_{mt} \propto \frac{\Delta\hat{c}_1(0,s)}{\Delta\hat{J}_1(0,s)} = \frac{a_{12}}{a_{22}} = \frac{\Lambda\tanh(\lambda u)+\tanh(u)}{mu[\Lambda\tanh(u)\tanh(\lambda u)+1]} \tag{10}$$

By putting complex number s = jω in the above equation (10), the net complex impedance due to bilayer diffusion can be obtained. This expression is incorporated into the ZSimpwin framework as the J symbol. Thus, the proposed equivalent circuit model for the electrochemical cell is represented below in Figure 7(e). Small perturbation AC impedance spectrum for the cell, cycled for 100 cycles, is presented in Figure 7(f). The proposed equivalent circuit fits well with the experimental values. The fit parameters are presented in the supporting information. From the values obtained post fitting, it is observed that bilayer diffusion occurs with the thickness of the passivation layer seemingly smaller than the thickness of the SEI Layer (for the superconcentrated electrolyte case). This can be seen by calculating the ($\lambda/\Lambda$) value, which is the ratio of the thickness of the two layers ($L_2/L_1$). A thinner passivation layer doesnot restrict the migration of the Li-ions through it. In some cases, it still allows an electrode to operate as a quasi-reference electrode, which is very much needed for two-electrode based systems such as a coin-cell electrochemical testing.[45] On the other hand, it is observed that this passivation layer is thicker when relatively dilute electrolytes are used (Supporting Information). Furthermore, a thick passivation layer can lead to a drastic voltage drop at the lithium foil surface, which is the reference electrode in a half-cell configuration. This will invariably lead to a voltage fading during cycling. This fading is in fact manifested due to this passivation layer, rather than from the material itself. The diffusion time constant ($L_1/D_1^{1/2}$), for the passivation layer, is smaller in the case of superconcentrated electrolyte compared to relatively dilute electrolytes. This is because the $\omega_o$ value is higher for case of passivation layer formed from superconcentrated electrolyte. Another factor which is indirectly speculated is that the diffusivity of lithium ions is high in the SEI ($D_2$) formed in the case of superconcentrated electrolyte ($\Lambda=(D_1/D_2)^{1/2}$). Formation of a highly compact SEI from superconcentrated electrolyte with good diffusional properties has also been reported in the literature.[46, 47] Thus, the characteristics of the unavoidably formed passivation layer over Li foil are favorable when super-concentrated electrolytes are used. Super-concentrated electrolytes are also found to be beneficial in retarding the vanadium dissolution in Kazakhstanite phase *via* the solubility limit approach, thereby improving the cycleability of the Kazakhstanite phase.

## 4. CONCLUSION

In conclusion, Kazakhstanite phase based Fe-V-O layered oxide has been successfully investigated for its potential use as a cathode material for the Li-ion batteries. The layered Kazakhstanite exhibits a high specific capacity of ~300mAhg$^{-1}$ between 1.5V-3.8V wrt. Li$^+$/Li redox couple, and thus, making it suitable for use as a cathode material for the Li-ion batteries. The Kazakhstanite phase is synthesized in a cost-efficient manner, without the use of any heat treatment procedures. One inherent disadvantage of this material is that it discharges first, instead of charging. This property is expected to limit its viability, since lithiated anodes must be used in-order to construct full cells.[48] Another disadvantage of this electrode material is that it exhibits poor cycling behavior with commercially available compositions of electrolyte. The concept of solubility limit for arresting active material dissolution in layered Kazakhstanite phase is demonstrated in order to improve its cycling performance. Super-concentrated electrolytes, with a high salt concentration of ~7M arrests the active material dissolution by creating a solubility limit for the dissolving species from the active material, owing to the low concentration of free solvent molecules in the electrolyte. Using a super-concentrated electrolyte affects the rate capability of the material. Though, it is observed that the material achieves close to 85 % of its maximum observed specific capacity (250mAhg$^{-1}$ out of 300mAhg$^{-1}$) in an hour, and deliver the charge at a constant specific current of 100mAg$^{-1}$ with high coulombic efficiency (~98.5%). The dissolved species from the active material is observed to migrate towards lithium counter electrode and deposit over it as a passivation layer. The lithium diffusion through this passivation layer follows linear bounded diffusion process. The characteristics of the passivation layers, formed from superconcentrated electrolytes, are experimentally validated to be superior in comparison to those formed from relatively dilute electrolytes. The Fe/V passivation layer is thinner in case of superconcentrated electrolyte, which results in a lower impedance towards $Li^+$ diffusion through it. Thus, in this part I, we are successful in demonstrating the benefits of using superconcentrated electrolytes to arrest vanadium dissolution in vanadium containing cathode materials. A coherent reasoning based on the improvement of the passivation layer characteristics over the Li foil surface is proposed in this part. With the improvement in the cycleability of the Kazakhstanite phase using super-concentrated electrolytes, high energy density lithium-ion batteries with long term cycling capability can be constructed which may outperform the existing technologies in the market. The proposed solubility limit approach is hypothesized to a generalized strategy, which is not material specific.

## SUPPORTING INFORMATION

The supporting information file contains calculations on the electron diffraction patterns, thermogravimetric analysis results, x-ray diffraction data from Indus-2 beamline, proposed unit cell structure, KFM topography and surface potential maps, ex-situ XRD and XPS results, preliminary electrochemical properties with relatively dilute electrolyte compositions and fitting parameters from EIS results, full cell electrochemical properties with lithiated graphite.

## AUTHOR INFORMATION

**Author Contributions**

The manuscript was prepared through contributions of all authors (experimentation, analysis and writing). All authors have given approval to the final version of the manuscript.

**Funding Sources**

S.D. and S.B.M. would like to acknowledge the funding from MHRD for project code UAY_IITKGP_001 (UAY Phase II), with approval order number F.No. 35-8/2017-TS.1, Dt: 05-12-2017, to carry out few of the parts of the work. S.D. would like to acknowledge the partial financial support from MHRD vide sanction F.16-59/2011-TEL, dated 30-09-2011 to carry out few parts of the work. S.B.M. would like to acknowledge partial financial support obtained from SERB, DST vide sanction EMR/2016/007537, dated 16-03-2018 and SGCIR grant of IIT Kharagpur vide approval No. IIT/SRIC/MS/MPV_ICG_2017_SGCIR/2018-19/080 dated 15-07-2018 to carry out few parts of the work. A.M. would like to acknowledge the funding received to carry out the research under the Prime Minister Research Fellowship (PMRF) scheme.

## ACKNOWLEDGMENT

A.M. would like to acknowledge MHRD, Govt. of India for the Prime Minister Research Fellowship. S.J would like to acknowledge MHRD, Govt. of India for the Ph.D Fellowship. The authors would like to thank Dr. A.K. Sinha and his team for X-ray Diffraction experiment at BL-12 Beamline at Indus 2, RRCAT. The authors would like to acknowledge BL-14 Beamline at Indus-2 for XPS experiments. The authors would like to acknowledge the FESEM facility sponsored by DST-FIST at Materials Science Centre, Indian Institute of Technology Kharagpur for scanning electron microscopy experiments. The authors would like to acknowledge the FESEM facility sponsored by DST-FIST at Department of Metallurgical and Materials Engineering, Indian Institute of Technology Kharagpur for scanning electron microscopy and energy dispersive x-ray spectroscopy experiments.

**Figures:**

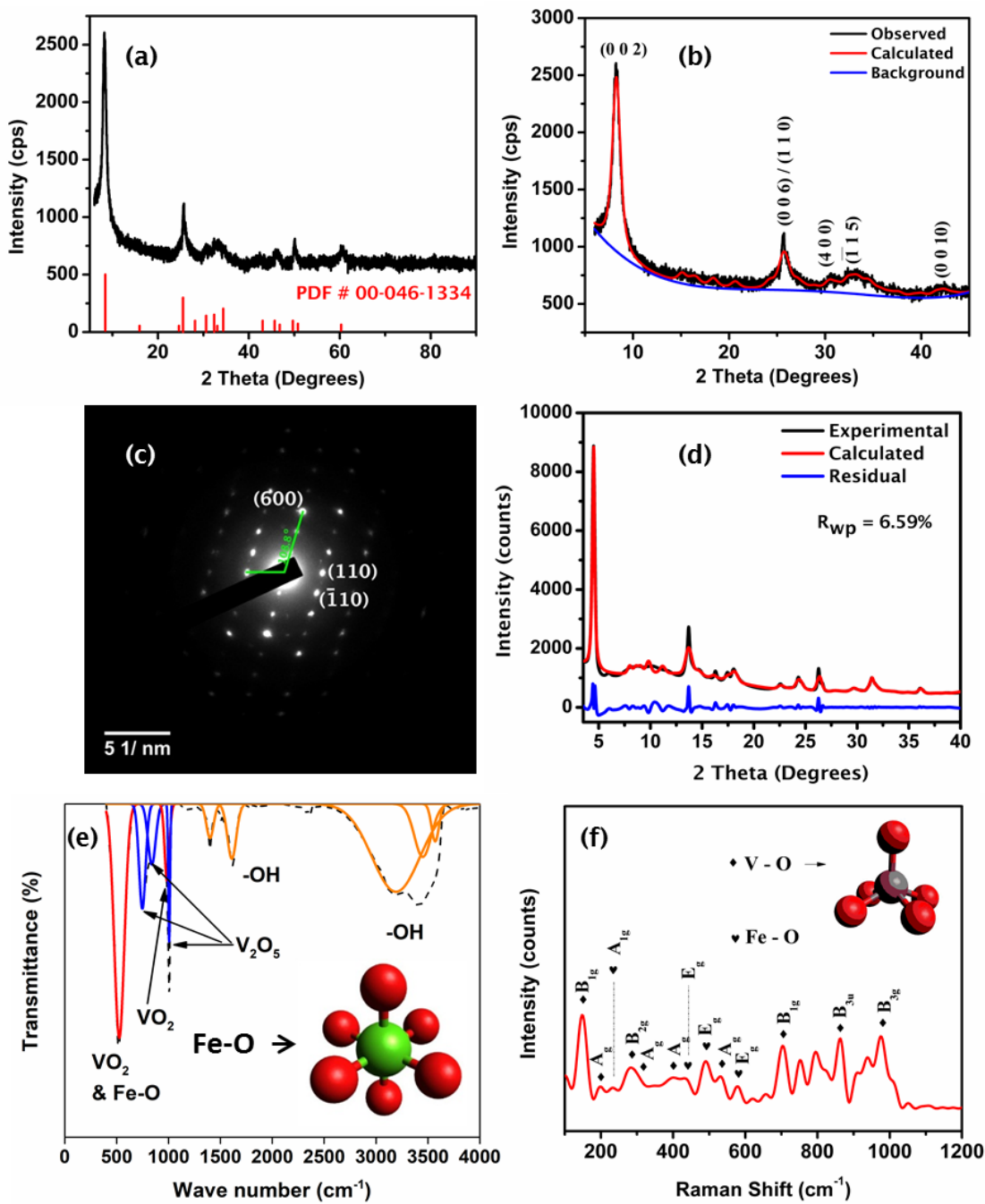

**Figure 1. (a)** X-ray diffractogram of the as-prepared powders, whose peaks are matched with the ICDD PDF Card for Kazakhstanite phase (00-046-1334). **(b)** Pawley refinement of the x-ray diffractogram for confirmation of phase matching and indexing. **(c)** Selected area diffraction

pattern for as-prepared powder along ZA (001). **(d)** Rietveld refined x-ray diffractogram with the proposed unit cell model of Kazakhstanite phase. **(e)** Fourier-Transform Infrared spectrum of as-prepared powder, with the co-ordination geometry of Fe-O bonds represented inset. **(f)** Raman spectrum of as-prepared powder, with the co-ordination geometry of V-O bonds represented inset.

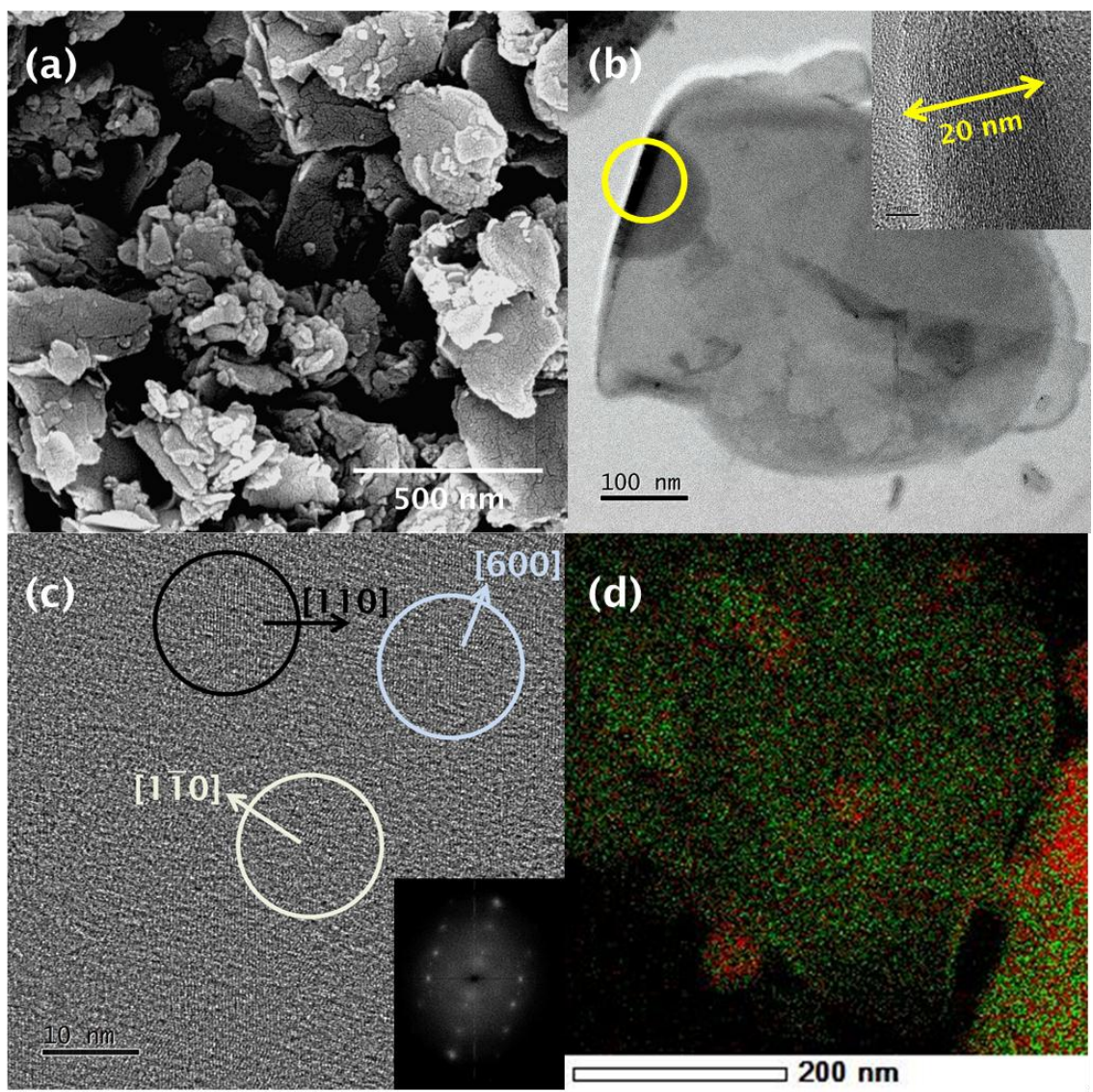

**Figure 2: (a)** Scanning electron micrograph of as-prepared powder, indicating the flake type morphology of the particles. **(b)** Transmission electron micrograph of one isolated particle,

indicating the width of the flake to be about 20nm (inset). **(c)** HRTEM image in the bulk region of the particle shown in (b), indicating the various lattice fringes and their directions along with the FFT spectrum in inset. **(d)** EDS overlay map of Fe and V, indicating a uniform distribution of the two elements in the as-prepared powder.

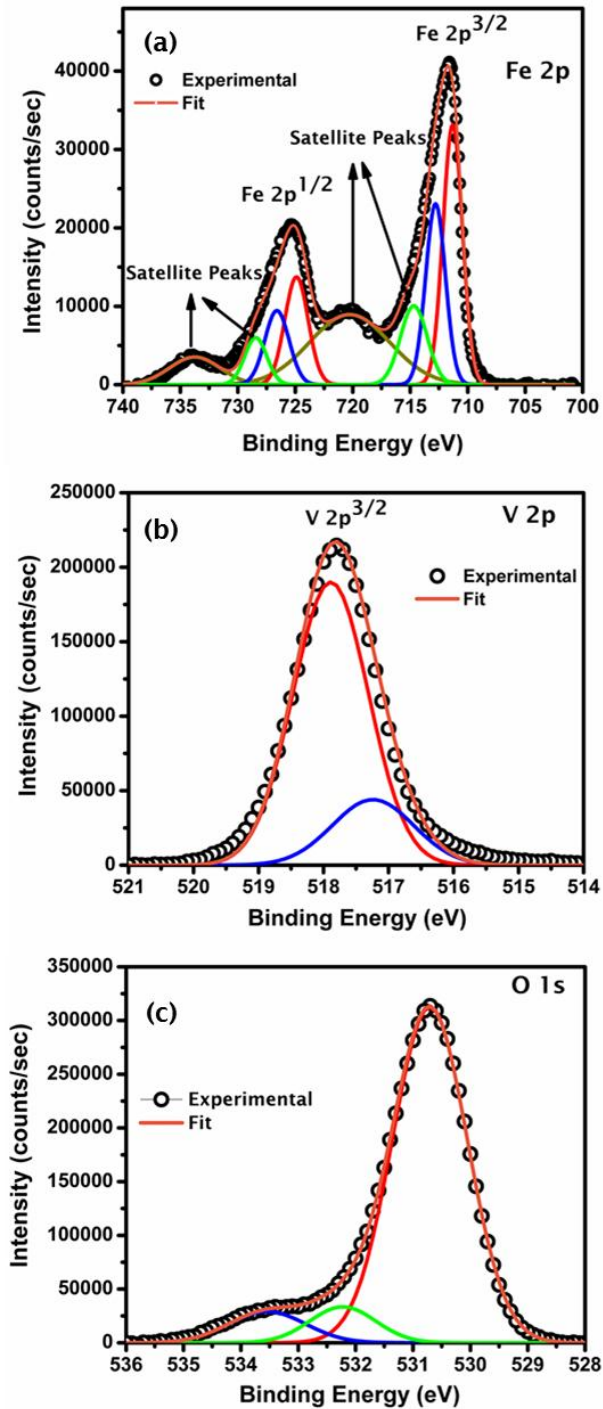

**Figure 3:** Core level XPS spectra for **(a)** Fe 2p, **(b)** V 2p, and **(c)** O 1s, indicating the various oxidation states of the elements present in the as-prepared powders along with their peak deconvolution.

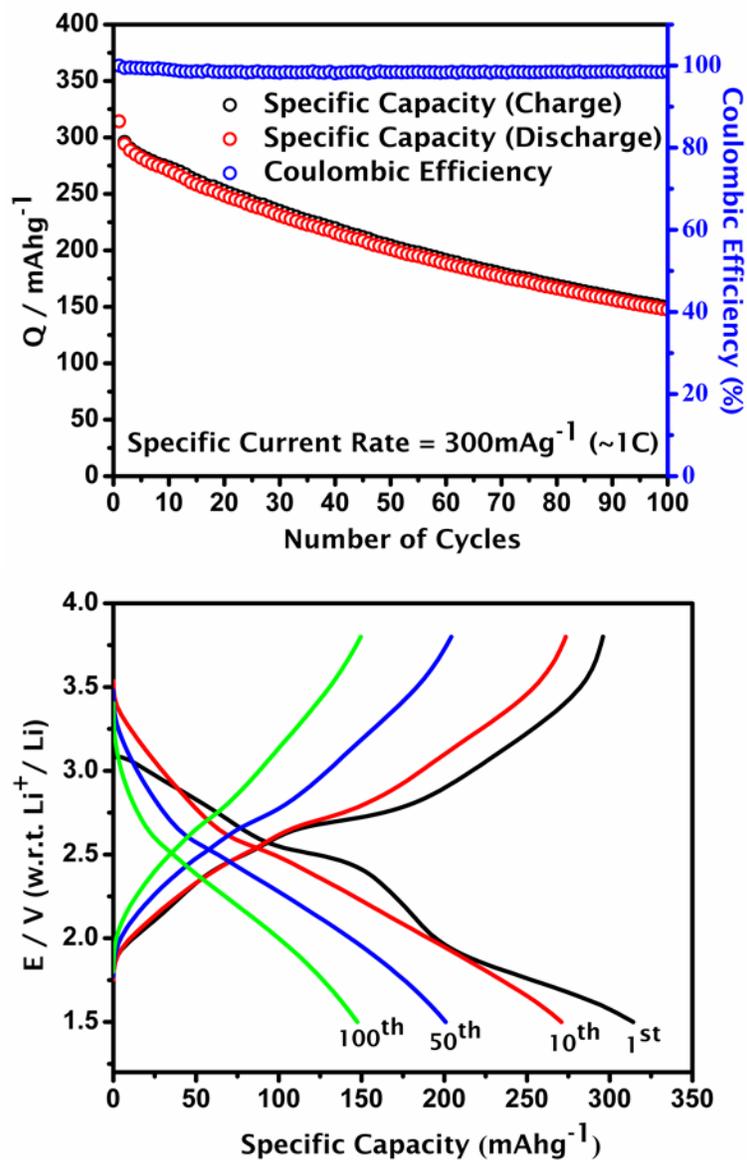

**Figure 4: (a)** Cycleability plot of the electrodes, for 100 cycles, tested with EE electrolyte, at a specific current of 300mAg⁻¹. **(b)** Charge-Discharge profile of the electrodes recorded between 1.5V and 3.8V at various cycles, mentioned in the plot.

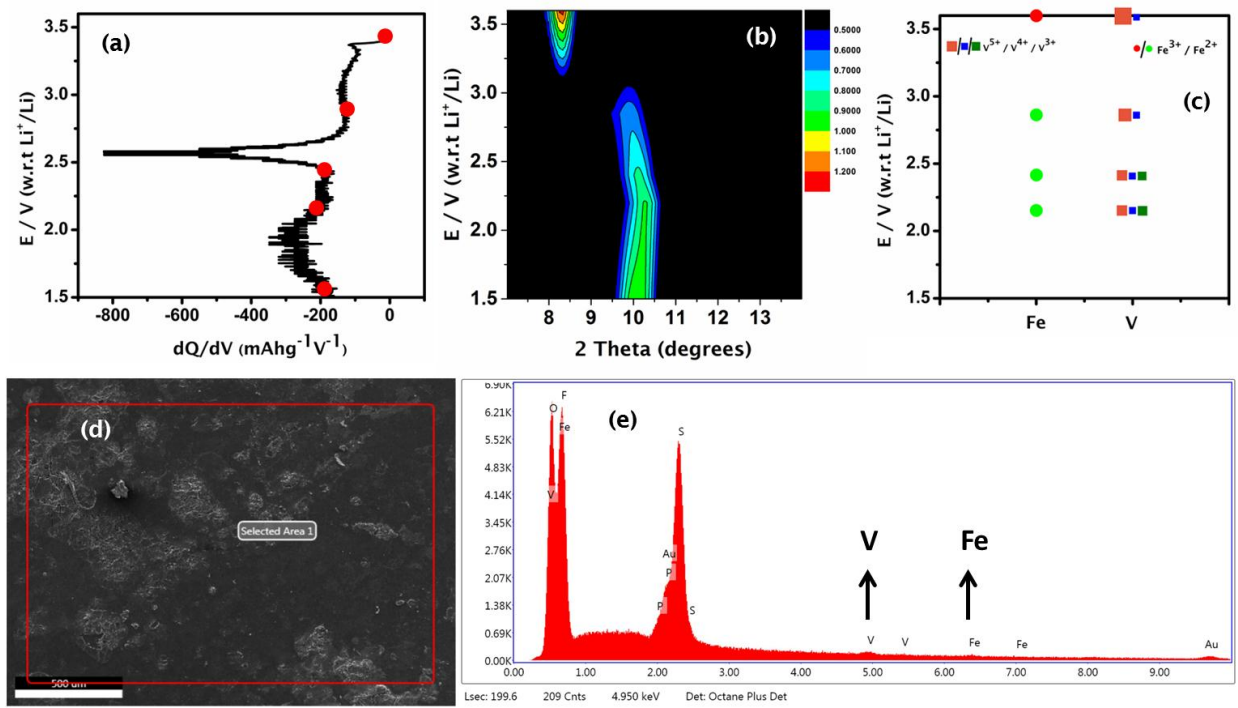

**Figure 5: (a)** Derivative plot for the charge-discharge profile, for identification of regions where major structural or electrochemical reactions may have taken place (red spots indicating points of interest). **(b)** Contour plot of x-ray diffractograms for (002) peak at different states of charge (SOC), indicating the initial contraction of the lattice along [001] direction during lithiation from OCP to 2.2V, followed by slight increase in the d-spacing for SOC below 2.2V. **(c)** Oxidation change map from ex-situ XPS results indicating the various oxidation states of Fe and V at different SOC. The color of the symbol is synonymous to the color of the compounds of Fe and V found in those oxidation states naturally, for easy representation. The size of the symbol represents an approximation towards the fraction of the oxidation states present. **(d)** Scanning electron micrograph of the area of interest of the cycled separator (side facing the lithium foil) for EDS analysis. **(e)** EDS spectrum of the cycled separator, indicating the presence of Fe and V on it.

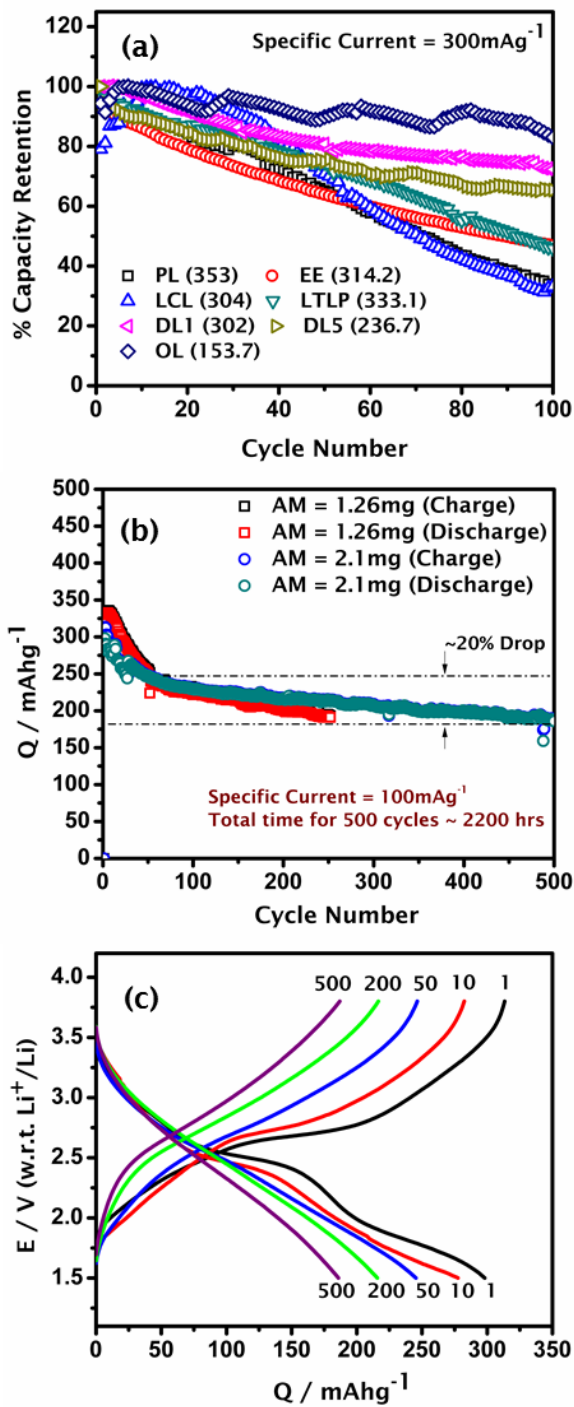

**Figure 6: (a)** Normalized cycleability plots for the electrodes tested with the different electrolytes mentioned in Table 1, for 100 cycles. The specific current for testing is 300mAg⁻¹, and the normalization factors are bracketed. **(b)** Cycleability plot for the electrodes, with different active material loading, tested with OL electrolyte at a specific current of $100mAg^{-1}$. **(c)** Charge-Discharge profile at different cycles, tested with OL electrolyte, for the electrode containing 2.1mg active material loading.

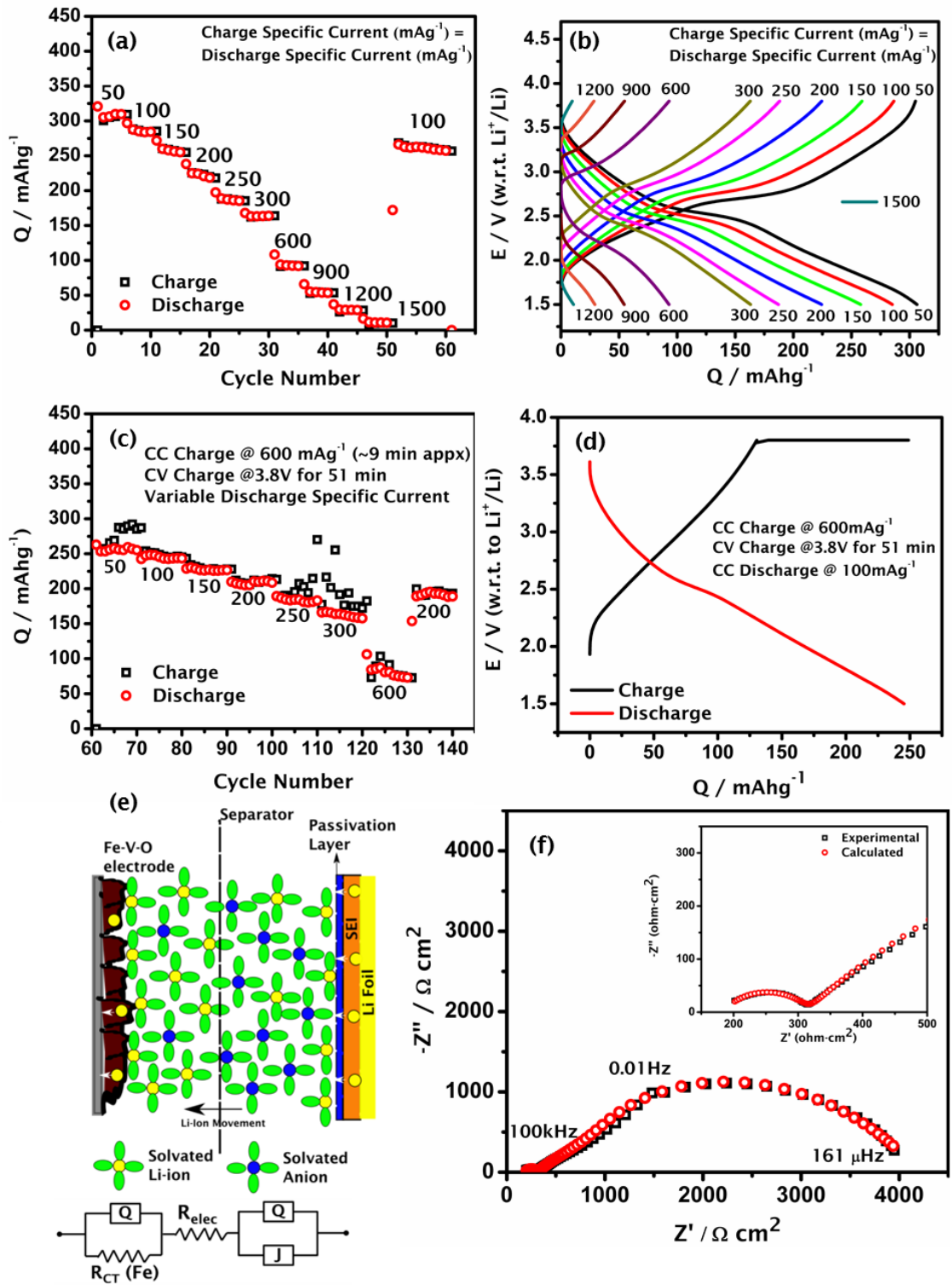

**Figure 7: (a)** Rate capability plots for the electrodes tested between 1.5V-3.8V at different specific current ranging from 50-1500mAg⁻¹. **(b)** Charge-Discharge profiles for the electrodes

obtained at different specific current. **(c)** Rate capability plots for the same electrode tested with a CCCV charging scheme (CC charging @600mAg$^{-1}$, CV charge @3.8V for 51 min) and variable discharge specific current. **(d)** Charge-Discharge profile for the same electrode tested with the above mentioned CCCV scheme and a discharge specific current of 100mAg$^{-1}$. **(e)** Proposed model of the electrochemical cell with OL electrolyte with the equivalent circuit represented below. **(f)** Nyquist spectrum collected for the coin cell after 100 cycles, with the high frequency region represented inset.